\documentclass[twocolumn,showpacs,aps,prl,superscriptaddress]{revtex4}

\usepackage{graphicx}
\usepackage{dcolumn}
\usepackage{amsmath}
\usepackage{epsfig}
\usepackage{feynmp}
\RequirePackage{xspace}

\usepackage{relsize}
\def\babar{\mbox{\slshape B\kern-0.1em{\smaller A}\kern-0.1em
    B\kern-0.1em{\smaller A\kern-0.2em R}}}

\def\t     {\ensuremath{t}\xspace}

\def\pim   {\ensuremath{\pi^-}\xspace}

\def\Kbar  {\kern 0.2em\overline{\kern -0.2em K}{}\xspace}

\def\Kz    {\ensuremath{K^0}\xspace}
\def\Kzb   {\ensuremath{\Kbar^0}\xspace}
\def\KzKzb {\ensuremath{\Kz \kern -0.16em \Kzb}\xspace}
\def\Kp    {\ensuremath{K^+}\xspace}
\def\Km    {\ensuremath{K^-}\xspace}

\def\KpKm  {\ensuremath{\Kp \kern -0.16em \Km}\xspace}

\def\Kstarz  {\ensuremath{K^{*0}}\xspace}

\def\Dbar    {\kern 0.2em\overline{\kern -0.2em D}{}\xspace}

\def\Dz      {\ensuremath{D^0}\xspace}
\def\Dzb     {\ensuremath{\Dbar^0}\xspace}
\def\DzDzb   {\ensuremath{\Dz {\kern -0.16em \Dzb}}\xspace}
\def\Dp      {\ensuremath{D^+}\xspace}
\def\Dm      {\ensuremath{D^-}\xspace}

\def\DpDm    {\ensuremath{\Dp {\kern -0.16em \Dm}}\xspace}

\def\B       {\ensuremath{B}\xspace}
\def\Bbar    {\kern 0.18em\overline{\kern -0.18em B}{}\xspace}

\def\BB      {\ensuremath{B\Bbar}\xspace} 
\def\Bz      {\ensuremath{B^0}\xspace}
\def\Bzb     {\ensuremath{\Bbar^0}\xspace}
\def\BzBzb   {\ensuremath{\Bz {\kern -0.16em \Bzb}}\xspace}
\def\Bu      {\ensuremath{B^+}\xspace}
\def\Bub     {\ensuremath{B^-}\xspace}

\def\Bm      {\ensuremath{\Bub}\xspace}

\def\BpBm    {\ensuremath{\Bu {\kern -0.16em \Bub}}\xspace}

\def\BorBbar    {\kern 0.18em\optbar{\kern -0.18em B}{}\xspace}
\def\DorDbar    {\kern 0.18em\optbar{\kern -0.18em D}{}\xspace}
\def\KorKbar    {\kern 0.18em\optbar{\kern -0.18em K}{}\xspace}

\def\jpsi     {\ensuremath{{J\mskip -3mu/\mskip -2mu\psi\mskip 2mu}}\xspace}

\mathchardef\Upsilon="7107
\def\Y#1S{\ensuremath{\Upsilon{(#1S)}}\xspace}% no space before {...}!

\def\FourS {\Y4S}

\mathchardef\Deltares="7101
\mathchardef\Xi="7104
\mathchardef\Lambda="7103
\mathchardef\Sigma="7106
\mathchardef\Omega="710A

\def\Deltabar{\kern 0.25em\overline{\kern -0.25em \Deltares}{}\xspace}
\def\Lbar{\kern 0.2em\overline{\kern -0.2em\Lambda\kern 0.05em}\kern-0.05em{}\xspace}
\def\Sigbar{\kern 0.2em\overline{\kern -0.2em \Sigma}{}\xspace}
\def\Xibar{\kern 0.2em\overline{\kern -0.2em \Xi}{}\xspace}
\def\Obar{\kern 0.2em\overline{\kern -0.2em \Omega}{}\xspace}
\def\Nbar{\kern 0.2em\overline{\kern -0.2em N}{}\xspace}
\def\Xb{\kern 0.2em\overline{\kern -0.2em X}{}\xspace}

\def\BR         {{\ensuremath{\cal B}\xspace}}

\def\mes        {\mbox{$m_{\rm ES}$}\xspace}

\def\DeltaE     {\mbox{$\Delta E$}\xspace}

\newcommand{\tev}{\ensuremath{\mathrm{\,Te\kern -0.1em V}}\xspace}
\newcommand{\gev}{\ensuremath{\mathrm{\,Ge\kern -0.1em V}}\xspace}
\newcommand{\mev}{\ensuremath{\mathrm{\,Me\kern -0.1em V}}\xspace}
\newcommand{\kev}{\ensuremath{\mathrm{\,ke\kern -0.1em V}}\xspace}
\newcommand{\ev}{\ensuremath{\mathrm{\,e\kern -0.1em V}}\xspace}
\newcommand{\gevc}{\ensuremath{{\mathrm{\,Ge\kern -0.1em V\!/}c}}\xspace}
\newcommand{\mevc}{\ensuremath{{\mathrm{\,Me\kern -0.1em V\!/}c}}\xspace}
\newcommand{\gevcc}{\ensuremath{{\mathrm{\,Ge\kern -0.1em V\!/}c^2}}\xspace}
\newcommand{\mevcc}{\ensuremath{{\mathrm{\,Me\kern -0.1em V\!/}c^2}}\xspace}
\def\invfb   {\ensuremath{\mbox{\,fb}^{-1}}\xspace}

\def\mus  {\ensuremath{\rm \,\mus}\xspace}

\def\mus        {\ensuremath{\,\mu{\rm s}}\xspace}    %% microsecond
\def\to                 {\ensuremath{\rightarrow}\xspace}

\def\pep2{PEP-II}

\def\gsim{{~\raise.15em\hbox{$>$}\kern-.85em
          \lower.35em\hbox{$\sim$}~}\xspace}
\def\lsim{{~\raise.15em\hbox{$<$}\kern-.85em
          \lower.35em\hbox{$\sim$}~}\xspace}

\def\CP                {\ensuremath{C\!P}\xspace}
\def\deltaz{\ensuremath{{\rm \Delta}z}\xspace}
\def\deltat{\ensuremath{{\rm \Delta}t}\xspace}
\def\deltamd{\ensuremath{{\rm \Delta}m_d}\xspace}

\xspace

\newcommand{\jprlBase}       {Phys.\ Rev.\ Lett.\xspace}
\newcommand{\jprBase}        {Phys.\ Rev.\xspace}
\newcommand{\jplBase}        {Phys.\ Lett.\xspace}

\newcommand{\nimBaseC}       {Nucl.\ Instr.\ and Methods\xspace}

\newcommand{\zpBase}         {Z.\ Phys.\xspace}

\newcommand{\nim}       [1]  {\nimBaseC~{\bf #1}}
\newcommand{\plb}       [1]  {\jplBase\ B~{\bf #1}}

\newcommand{\jprl}      [1]  {\jprlBase\ {\bf #1}}
\newcommand{\pr}        [1]  {\jprBase\ {\bf #1}}
\newcommand{\jprd}      [1]  {\jprBase\ D~{\bf #1}}

\newcommand{\progtp}    [1]  {{Prog.\ Th.\ Phys.\ {\bf #1}}}

\newcommand{\zp}        [1]  {\zpBase\ {\bf #1}}
\def\jetset74   {\mbox{\tt Jetset \hspace{-0.5em}7.\hspace{-0.2em}4}\xspace}
\newcommand{\stwobg} {\ensuremath{\sin(2\beta+\gamma)}}
\def\myprl  #1 #2 #3 {\jprl{#1},\ #2 (#3)}
\def\myplb  #1 #2 #3 {\plb{#1},\ #2 (#3)}
\def\myprd  #1 #2 #3 {\jprd{#1},\ #2 (#3)}
\def\mynim  #1 #2 #3 {\nim{#1},\ #2 (#3)}
\def\mypr   #1 #2 #3 {\pr{#1},\ #2 (#3)}
\def\sbglim   {0.69}
\def\figurebox#1#2#3{%

    \def\arg{#3}%

    \ifx\arg\empty

    {\hfill\vbox{\hsize#2\hrule\hbox to #2{\vrule\hfill\vbox to #1{\hsize#2\vfill}\vrule}\hrule}\hfill}%

    \else

    {\hfill\epsfbox{#3}\hfill}%

    \fi}

\begin{document}
%\preprint{\babar-PUB-\BABARPubYear/\BABARPubNumber} 
%\preprint{SLAC-PUB-\SLACPubNumber} 
%\noindent
%\babar-PUB-\BABARPubYear/\BABARPubNumber\\
%SLAC-PUB-\SLACPubNumber\\
%hep-ex/\LANLNumber\vskip 0.4cm

\title{
{\large \bf Measurement of time-dependent \CP asymmetries in $\Bz{\to} D^{(*)\pm}\pi^{\mp}$ decays and constraints on $\stwobg$}
}

% Input author list file
%\input pubboard/authors_aug2003.tex
%% author list as of 08-Aug-2003 (601 authors)
%
\author{B.~Aubert}
\author{R.~Barate}
\author{D.~Boutigny}
\author{J.-M.~Gaillard}
\author{A.~Hicheur}
\author{Y.~Karyotakis}
\author{J.~P.~Lees}
\author{P.~Robbe}
\author{V.~Tisserand}
\author{A.~Zghiche}
\affiliation{Laboratoire de Physique des Particules, F-74941 Annecy-le-Vieux, France }
% ...
\collaboration{The \babar\ Collaboration}
\noaffiliation

\date{August $27^{th}$, 2003}

% Abstract
\begin{abstract}
We present a  measurement of  \CP-violating asymmetries in 
fully reconstructed $\Bz{\to}D^{(*)\pm}\pi^{\mp}$ decays in 
 approximately $88$ million \Y4S $\to$ \BB decays collected with the
\babar\ detector at the PEP-II asymmetric-energy $B$ factory at SLAC.
From a time-dependent maximum likelihood fit we obtain for the \CP-violating parameters:
$a = -0.022\pm0.038 \,(\textrm{stat.})\pm 0.020 \,(\textrm{syst.}),   
a^{*} = -0.068\pm0.038  \,(\textrm{stat.})\pm 0.020 \,(\textrm{syst.}), 
c_{\rm lep} = +0.025\pm0.068  \,(\textrm{stat.})\pm 0.033 \,(\textrm{syst.})$,  and
$c^{*}_{\rm lep} = +0.031\pm0.070  \,(\textrm{stat.})\pm 0.033 \,(\textrm{syst.}). $
Using other measurements and theoretical assumptions we interpret the results in terms of the angles of the
Cabibbo-Kobayashi-Maskawa unitarity triangle, and
find $|\stwobg|>\sbglim$ at $68\%$ confidence
level. We exclude the hypothesis of no \CP violation ($\stwobg=0$) at $83\%$ confidence level.
\end{abstract}

% PACS, the Physics and Astronomy Classification Scheme.

\maketitle

In the Standard Model, \CP\ violation in the weak interactions between quarks
manifests itself as a non-zero area of the Cabibbo-Kobayashi-Maskawa (CKM) unitarity triangle~\cite{CKM}.
While it is sufficient to measure one of its angles $\alpha$, $\beta$, or
$\gamma$ to be different from $0$ or $180^{\circ}$ to demonstrate the
existence of \CP\ violation, the unitarity triangle needs to be overconstrained with different measurements
to  test the CKM mechanism.
Measurements of $\beta$ free from theoretical uncertainties exist~\cite{babar_sin2b,belle_sin2b}, but there are no such
measurements of  $\alpha$ and $\gamma$. 
This letter reports the measurement of \CP-violating asymmetries
in  $\Bz{\to} D^{(*)\pm} \pi^{\mp}$ decays~\cite{chconj} in $\FourS\to \BB$
decays and its interpretation in terms of
constraints on $|\sin(2\beta+\gamma)|$~\cite{sin2bg,fleischer}.

The time evolution of $B^0{\to} D^{(*)\pm} \pi^{\mp}$ decays is
sensitive to $\gamma$ because of the interference between the CKM-favored 
decay $\Bzb{\to} D^{(*)+} \pi^-$, whose amplitude is proportional to the CKM matrix
elements $V^{}_{cb}V^*_{ud}$, and the doubly-CKM-suppressed  
decay $\Bz{\to} D^{(*)+}\pi^-$, whose amplitude is proportional to $V^{}_{cd}V^*_{ub}$. % (Fig.~\ref{fig:feyn}). 
The relative weak phase between the two  amplitudes is $\gamma$,
which, when combined with $\BzBzb$ mixing, yields
a weak phase difference of $2\beta+\gamma$ between the interfering amplitudes.

The decay rate distribution for $B^0{\to} D^{\pm}\pi^{\mp}$ decays is
\begin{eqnarray}
f^{\pm}(\eta,\deltat) &=& \frac{e^{-\left|\deltat\right|/\tau}}{4\tau} \times \\
 &&[1 \mp S_\zeta \sin(\deltamd\deltat) \mp\eta C \cos(\deltamd\deltat)]\,, \nonumber
\label{eq:fplus}
\end{eqnarray}
where $\tau$ is the \Bz lifetime, neglecting the decay width difference, $\deltamd$ is the $\Bz\Bzb$ mixing frequency,
and $\deltat = t_{\rm rec} - \t_{\rm tag}$ is the time of the  
$B^0\to D^{\pm} \pi^{\mp}$ decay ($B_{\rm rec}$) relative to the decay of the other $B$ ($B_{\rm tag}$). 
In this equation the upper (lower) sign refers to the flavor of $B_{\rm tag}$ as \Bz(\Bzb),
while $\eta=+1$ ($-1$) and $\zeta=+$ ($-$) for the final state $D^{-}\pi^{+}$ ($D^{+}\pi^{-}$).
In the Standard Model, the $S$ and $C$ parameters can be expressed as 
% The $S$ and $C$ parameters can be expressed as 
%
\begin{eqnarray}
S_\pm = -\frac{2\textrm{Im}(\lambda_\pm)}{1+|\lambda_\pm|^2}\,, \hspace{0.4cm} {\rm and} 
\hspace{0.4cm}
C=\frac{1-r^2}{1+r^2}\,,
\label{eq:cands}
\end{eqnarray}
%
%
%where we define $r \equiv |\lambda_+|$ = 
%$1/|\lambda_-|$, $\lambda_\pm= (q/p)
% A(\Bzb{\to} D^{\mp}\pi^\pm)/A(\Bz{\to}
%D^{\mp}\pi^\pm)=r^{\pm 1}e^{-i(2\beta+\gamma\mp\delta)}$. Here $q/p$ is a function of the elements of the mixing
%matrix~\cite{PDG2002} and $\delta$ is the relative strong phase between the two contributing amplitudes.
%
where  $\lambda_\pm= r^{\pm 1}e^{-i(2\beta+\gamma\mp\delta)}$. Here
$\delta$ is the relative strong phase and $r$ is the magnitude of the 
ratio of the suppressed and the favored amplitudes.
The same equations apply for $B^0{\to} D^{*\pm} \pi^{\mp}$ decays, with 
$r$ and $\delta$ replaced by the parameters $r^*$ and
$\delta^*$, respectively~\cite{note}.

%analysis strategy
The analysis strategy is similar to that of the time-dependent mixing 
measurement performed at \babar~\cite{sin2b_prd}.
To identify the flavor of $B_{\rm tag}$, each event
is assigned by a neural network to one of four hierarchical, mutually exclusive
tagging categories:
one lepton and two kaon categories based  on the charges of 
identified leptons and kaons, and a fourth category 
for remaining events.
The effective tagging efficiency is ($28.1\pm 
0.7$)\%~\cite{babar_sin2b}. The time
difference $\deltat$ is calculated from the  separation along the
beam collision axis, $\deltaz$, between the $B_{\rm rec}$ and $B_{\rm tag}$ decay 
vertices. We determine the  $B_{\rm rec}$ vertex from its charged tracks. 
The  $B_{\rm tag}$ decay vertex is obtained by fitting tracks that do not belong
to $B_{\rm rec}$, 
imposing constraints from the $B_{\rm rec}$ momentum and the beam-spot location. 
The $\deltat$ resolution is approximately $1.1$ ps.

The expected \CP\ asymmetry in these decays is small ($r^{(*)} \approx
|V_{ub}^{*}V^{}_{cd}/V_{ud}^{*}V^{}_{cb}|\approx 0.02$), and therefore 
this measurement is sensitive to 
 the interference between the $b\to u$
and  $b\to c$ amplitudes in the decay of $B_{\rm tag}$. To account for this effect 
we use a parametrization different from Eq.~\ref{eq:fplus},
which is described in Ref.~\cite{DCSD} and summarized here. 
%This effect depends on the $B_{\rm tag}$ decay modes.
For each tagging category ($i$) the interference is
parametrized in terms of the effective parameters 
$r^\prime_i$ and $\delta^\prime_i$.
Neglecting terms of order $r^{(*)2}$ and $r^{\prime 2}_i$, for each tagging category the 
\deltat
distribution becomes
\begin{eqnarray}
f^{\pm{(*)}}_i(\eta,\deltat) &=&
  \frac{e^{-\left|\deltat\right|/\tau}}{4\tau} \times [ 1 \mp (a^{(*)}
  \mp \eta b_i - \eta c_i^{(*)})\nonumber \\
  && \sin(\deltamd\deltat)\mp\eta\cos(\deltamd\deltat)]\,,
\label{completepdf}
\end{eqnarray}
where, in the Standard Model,
\begin{eqnarray}\nonumber
&a^{(*)}&=\ 2r^{(*)}\sin(2 \beta+\gamma)\cos\delta^{(*)}\,, \\ \nonumber
&b_i&=\ 2r^\prime_i\sin(2 \beta+\gamma)\cos\delta^\prime_i\,, \\
&c_i^{(*)}&=\ 2\cos(2 \beta+\gamma) (r^{(*)}\sin\delta^{(*)}-r^\prime_i\sin\delta^\prime_i)\,.
\label{acdep}
\end{eqnarray}
Semileptonic $B$ decays do not have a doubly-CKM-suppressed amplitude contribution, 
and hence $r^{\prime}_{\rm lep}=0$. 
Given that we have two $B$ decay modes and four tagging categories, we use two $a$ parameters (one for
each final state), three $b$ parameters (one for each non-lepton tagging
category), and eight $c$ parameters (one for each combination of
tagging category and final state).
Results are quoted only for the four parameters $a^{(*)}$ and
$c_{\rm lep}^{(*)}$, which are independent of the unknowns 
$r^\prime_i$ and $\delta^\prime_i$. The other parameters are allowed to float in
the fit, but, as they depend on $r^\prime_i$ and $\delta^\prime_i$, they do not 
contribute to the interpretation of the result in terms of \stwobg. 

% data sample
This measurement is based on $88$ million \Y4S $\to$ \BB decays, corresponding
to an integrated luminosity of $82\invfb$, collected 
with the \babar\ detector~\cite{detector} at the PEP-II asymmetric-energy $B$ factory at SLAC.
We use a Monte Carlo simulation of the \babar\ detector based on
GEANT4~\cite{geant} to validate the analysis procedure and to estimate
some of the backgrounds.

%%
%\begin{figure}[b]
%\begin{center}
%\epsfig{figure=max_1.eps,width=0.45\linewidth}\hspace{0.25cm}
%\epsfig{figure=max_2.eps,width=0.45\linewidth}
%\end{center}
%%\setlength{\abovecaptionskip}{1pt}
%\caption{Feynman diagrams for (a) $\bar{B^0}{\to} D^{(*)+}\pi^-$ and  (b) $\Bz{\to} D^{(*)+}\pi^-$. }
%\label{fig:feyn}
%\end{figure}
%%

The event selection and the reconstruction of $\Bz\to D^{(*)\pm}\pi^{\mp}$
candidates are detailed in Ref.~\cite{sin2b_prd}.
%
%The \Dstarp is reconstructed in its decay to $\Dz\pip$, where the \Dz subsequently decays to one of the four modes
%$K^{-}\pi^{+}$, $K^{-}\pi^{+}\piz$, $K^{-}\pi^{+}\pi^{-}\pi^{+}$, or $\KS\pi^{+}\pi^{-}$.
%
%The \Dp is reconstructed in its decays to $K^{-}\pi^{+}\pi^{+}$ and $\KS\pi^{+}$.
%
Signal and background are discriminated by two kinematic variables:
the beam-energy substituted mass, $\mes \equiv \sqrt{(\sqrt{s}/2)^{2} - {p_B^*}^2}$,
and the difference between the $B$ candidate's measured energy and the beam energy, 
$\DeltaE \equiv E_{B}^* - (\sqrt{s}/2)$, where $E_{B}^*$ ($p_B^*$) is the energy (momentum) of the \B\ candidate
in the $e^{+}e^{-}$ center-of-mass frame, and $\sqrt{s}$ is the total center-of-mass energy.
The signal region is defined as $|\DeltaE| <3\sigma$, where the resolution $\sigma$ is mode-dependent
and approximately $20\mev$, as determined from data. 
\begin{figure}[!tbp]
\begin{center}
\epsfig{figure=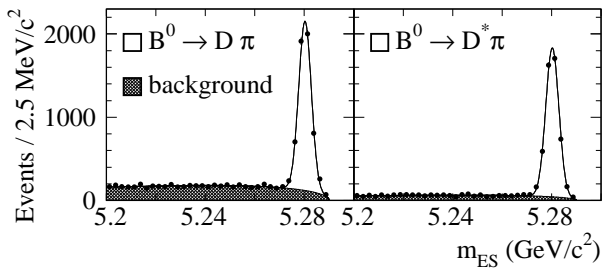,width=0.92\linewidth,clip=}

\end{center}
\caption{Distributions of \mes\ in the \DeltaE\ signal region for events with tagging 
information in the $\Bz
  \rightarrow D^{\pm}\pi^{\mp}$ (left plot) and the $B^0 \rightarrow 
D^{*\pm}\pi^{\mp}$
 sample (right plot).}
\label{mes}
\end{figure}
Figure~\ref{mes} shows the $\mes$ distribution for candidates in the
\DeltaE\ signal region. The \mes\ distribution is fit with the sum of a threshold function~\cite{Argus}, 
which accounts for the  background  from random
combinations of tracks, and a Gaussian distribution with a fitted width of about 2.5\mevcc\ describing the signal. 
After tagging, the Gaussian yield is $5207\pm 87$  and $4746\pm 78$
events for the $\Bz\to D^{\pm}\pi^{\mp}$  and 
$\Bz\to D^{*\pm}\pi^{\mp}$ sample respectively, with corresponding purities
 of $(84.9 \pm 0.5)\%$ and $(94.4 \pm 0.4)\%$ in a $\pm 3\sigma$ region around the nominal $B$ mass.
Backgrounds from $\Bz$ decays that peak in the $\mes$ signal region were estimated with Monte Carlo simulation to constitute 
$(0.21\pm 0.06)\%$ and $(0.13\pm 0.05)\%$ of  the
$\Bz\to D^{\pm}\pi^{\mp}$ and $\Bz\to D^{*\pm}\pi^{\mp}$
yields, respectively. 
For backgrounds from $B^{+}$  decays, the corresponding figures are $(0.93\pm 0.23)\%$ 
and $(0.93\pm 0.10)\%$.

An unbinned maximum-likelihood fit is performed on the selected $B$ candidates using the  \deltat\ distribution
in Eq.~\ref{completepdf}, convolved with a resolution function composed of three Gaussian distributions. 
Incorrect tagging dilutes the parameters $a^{(*)}$, $c_i^{(*)}$, and the coefficient of $\cos(\deltamd\deltat)$
by a factor $D_i=1-2w_i$~\cite{babar_sin2b,DCSD}, where $w_i$ is the mistag fraction.
The resolution function and the parameters associated with flavor tagging
are determined from the data and are consistent with previous \babar\ analyses~\cite{babar_sin2b}.
The combinatorial background is parametrized as the sum of a component
with zero lifetime and one with an effective lifetime fixed to the value obtained from simulation.
The fraction of each component and the 
 \deltat\ resolution parameters are 
left free in the fit to the data. The background coming from $B^{\pm}$ mesons
is modeled with an exponential decay with the  $B^{\pm}$ lifetime, and
its size is fixed to the value predicted by simulation.
The background from $B^{0}$
mesons is neglected in the nominal fit, but is considered in evaluating the systematic uncertainties.

\begin{figure*}
\begin{center}
\epsfig{figure=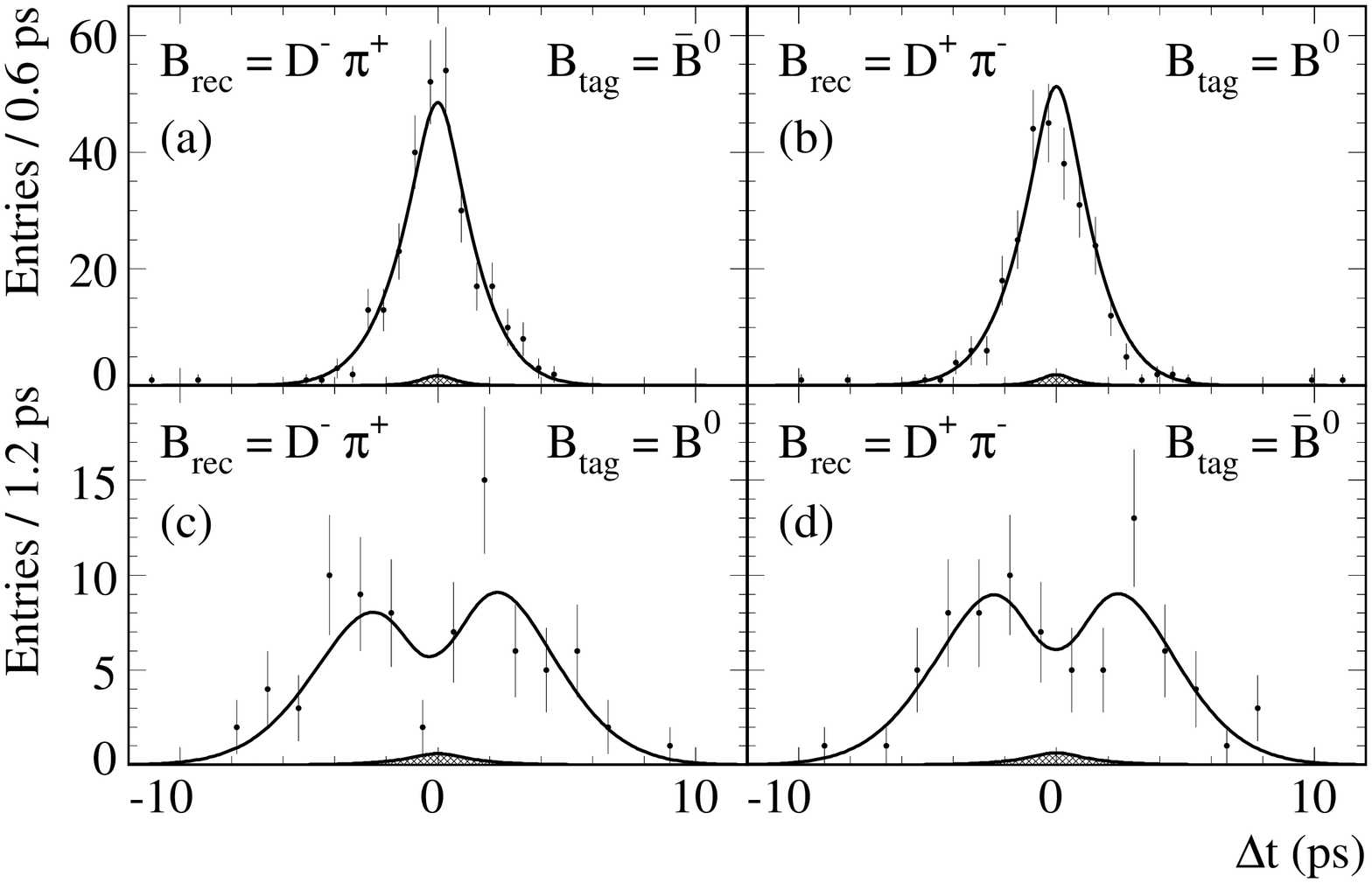,width=0.49\linewidth,clip=}
\epsfig{figure=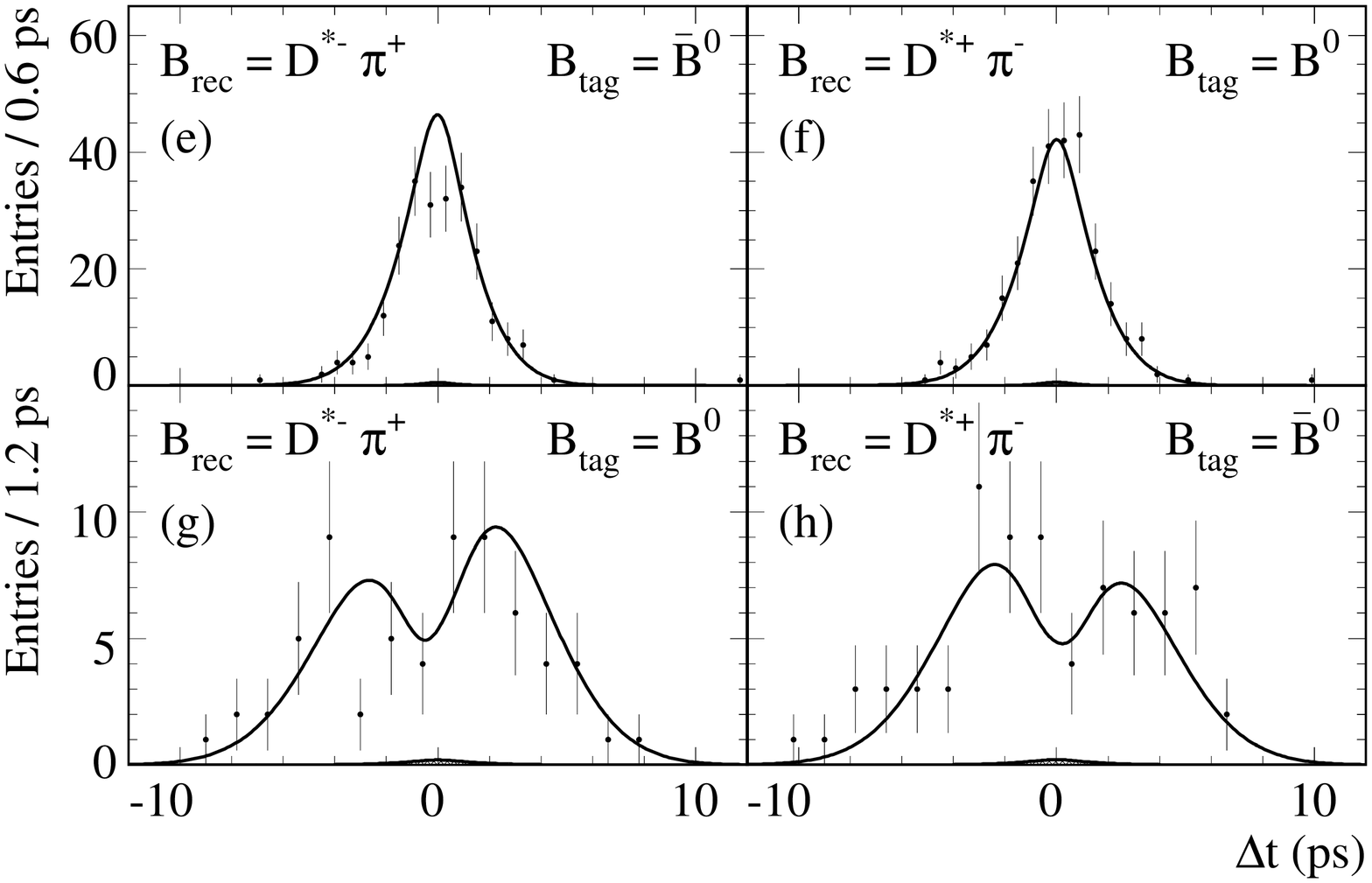,width=0.49\linewidth,clip=}
\end{center}
\caption{Distributions of \deltat\ for the  $\Bz{\to} D^{\pm}\pi^{\mp}$ (a-d) and $\Bz{\to} D^{*\pm}\pi^{\mp}$ (e-h) candidates tagged with leptons, 
split by  $B$ tagging flavor and  reconstructed final state. The lines are fit projections and hatched regions represent background.} 
\label{deltat1}
\end{figure*}

The results from the fit to the data are
\begin{eqnarray}\nonumber
a &=& -0.022\pm0.038 \,(\textrm{stat.})\pm 0.020 \,(\textrm{syst.})\,,  \\ \nonumber 
a^{*} &=& -0.068\pm0.038  \,(\textrm{stat.})\pm 0.020 \,(\textrm{syst.})\,, \\ \nonumber 
c_{\rm lep} &=& +0.025\pm0.068  \,(\textrm{stat.})\pm 0.033 \,(\textrm{syst.})\,,\\  
c^{*}_{\rm lep} &=& +0.031\pm0.070  \,(\textrm{stat.})\pm 0.033 \,(\textrm{syst.})\,. 
\label{acmeas}
\end{eqnarray}
All other fitted $b$ and $c$ parameters are consistent with zero. 
Figure~\ref{deltat1} shows the fitted $\deltat$ distributions for events from the lepton 
tagging category, which has the lowest level of background and mistag probability.
%
%\begin{figure}[!tbp]
%\begin{center}
%\epsfysize=6.5cm
%\epsfbox{dt6.eps}
%\end{center}
%\caption{Distributions of \deltat\ for the combined sample of $\Bz{\to} D^{*\pm}\pi^{\mp}$ candidates tagged with leptons, 
%split by the $B$ tagging flavor and the reconstructed final state. The lines are fit projections and hatched regions represent background.} 
%\label{deltat2}
%\end{figure}

The systematic uncertainties on the parameters in Eq.~\ref{acmeas}
has been calculated in a manner similar to that used in Ref.~\cite{sin2b_prd}.
A small bias in the $\deltat$ measurement could result
in a bias on  the  $c$ parameters in Eq.~\ref{completepdf}. 
For instance, a realistic $\deltat$ bias of $0.024$ ps results in a shift in
$c^{*}_{\rm lep}$ of
$0.002$. We are immune from this effect because we fit for
tagging category dependent biases in the resolution function 
directly on data.
Nonetheless, the impact of a possible mismeasurement of $\deltat$ 
has been estimated by varying the assumptions on the resolution function, 
the position of the beam-spot, the absolute $z$ scale, the internal
alignment of the vertex detector, and quality criteria on the
reconstructed vertex. The corresponding error on $a^{(*)}$ is
$\sigma_a=0.015$, while that on $c^{(*)}$ is  $\sigma_c=0.026$.
The systematic uncertainties on the fit technique ($\sigma_a=0.013$,
$\sigma_c=0.020$) include the upper limit on the fit bias estimated
from samples of fully simulated events, the uncertainty on the 
$B^{0}$ lifetime and \deltamd~\cite{PDG2002}, and the impact
of neglecting higher order terms in $r^{(*)}$ or $r^\prime_i$ in Eq.~\ref{completepdf}. 
As a cross-check, we performed the same fits on samples of 18233 $\Bm\to D^{(*)0}\pim$ and  
1740 $\Bzb\to \jpsi\Kstarz$ candidates,
where we find no significant \CP\ asymmetries, as expected.  
The systematic uncertainties in tagging ($\sigma_a=0.004$, $\sigma_c=0.003$) are estimated
allowing for different tagging efficiencies between \Bz\ and \Bzb\ and 
for different $\Delta t$ resolutions for correctly and incorrectly tagged events. 
We also account for uncertainties on the background ($\sigma_a=0.001$,
$\sigma_c=0.003$) by varying the effective lifetimes, dilutions, \mes\
shape parameters and signal fractions, 
and background \CP\ asymmetry up to five times the expected \CP\ asymmetry for signal.

The results can  be interpreted in terms of  $\stwobg$ (Eq.~\ref{acdep})
if the decay amplitude ratios $r^{(*)}$, expected to
be $|V_{ub}^{*}V^{}_{cd}/V_{ud}^{*}V^{}_{cb}|\approx 0.02$, are known.
Such small amplitude ratios cannot be determined from $\Bz{\to}
D^{(*)\pm}\pi^{\mp}$ events directly, because the current 
data sample is too small.
We estimate $r^{(*)}$ 
%from the ratios of branching fractions  $\BR(\Bz{\to} D_s^{(*)+}\pi^-)/{\BR(\Bz{\to}
%  D^{(*)-}\pi^+)}$~\cite{sin2bg} 
using the $SU(3)$ symmetry relation 
%\begin{equation}
$r^{(*)}= \tan\theta_{C}\sqrt{\frac{\BR(\Bz{\to} D_s^{(*)+}\pi^-)}{\BR(\Bz{\to} D^{(*)-}\pi^+)}}
\frac{f_{D^{(*)}}}{f_{{D_s}^{(*)}}}$~\cite{sin2bg}.
%\label{lambdadef}
%\end{equation}
%
From the measurements of the Cabibbo angle $\tan\theta_C$ = $0.2250\pm 0.0027$~\cite{PDG2002}, the branching fractions $\BR(\Bz{\to} D^{-}\pi^+)$ = $(0.30\pm0.04)\%$~\cite{PDG2002},  $\BR(\Bz{\to} D^{*-}\pi^+)$ = $(0.276\pm0.021)\%$~\cite{PDG2002}, $\BR(\Bz{\to} D_s^{+}\pi^-)$ = $(2.7_{-0.6}^{+0.7} \pm 0.8)\times 10^{-5}$~\cite{dspi}, $\BR(\Bz{\to} D_s^{*+}\pi^-)$ = $(1.9_{-1.3}^{+1.2}\pm0.5)\times 10^{-5}$~\cite{dspi} , and from calculations of the decay constant ratios 
$f_{D_s}/f_{{D}}$ = $1.11\pm0.01$ and  $f_{D_s^{*}}/f_{{D}^{*}}$ = $1.10\pm0.02$~\cite{fdsd} we obtain
%%all of which are listed in Table~\ref{tab:input}, we obtain
%
\begin{eqnarray}
r = 0.019 \pm 0.004\,,\hspace{1cm}
r^{*} = 0.017^{+0.005}_{-0.007}\,.
\label{lambdameas}
\end{eqnarray}
%
%We attribute an additional $30\%$ error to $r^{(*)}$ due to
%the unknown theoretical uncertainty on the validity of the $SU(3)$ symmetry assumption
%and to neglecting $W$-exchange contributions to $A(\Bz{\to} D^{(*)+}\pi^-)$.
%This error estimate is consistent with the spread in $r^{(*)}$ obtained using a variety of theoretical models~\cite{scr}.
%
%% CV
%%\begin{table}[ht] 
%%\begin{center}%\vspace{\baselineskip}
%%%\scriptsize
%%\caption{Input parameters for the calculation of $r^{(*)}$.}
%%\begin{ruledtabular}
%%\begin{tabular*}{\hsize}{l@{\extracolsep{0ptplus1fil}}c@{\extracolsep{0ptplus1fil}}r@{\extracolsep{0ptplus1fil}}}
%%%D{,}{\ \pm\|}{-1} }
%%Parameter & Value &\bridge Reference\\
%%\hline
%%$\tan\theta_C$ & $0.2250\pm 0.0027$ &\bridge \cite{PDG2002} \\
%%$\BR(\Bz{\to} D^{-}\pi^+)$ & $(0.30\pm0.04)\%$ &\bridge \cite{PDG2002} \\
%%$\BR(\Bz{\to} D^{*-}\pi^+)$ &  $(0.276\pm0.021)\%$ &\bridge \cite{PDG2002}\\
%%$\BR(\Bz{\to} D_s^{+}\pi^-)$ & $(2.7_{-0.6}^{+0.7} \pm 0.8)\times 10^{-5}$ &\bridge \cite{dspi}\\
%%$\BR(\Bz{\to} D_s^{*+}\pi^-)$ & $(1.9_{-1.3}^{+1.2}\pm0.5)\times 10^{-5}$ &\bridge \cite{dspi}\\
%%$f_{D_s}/f_{D}$ & $1.11\pm0.01$ &\bridge \cite{fdsd} \\
%%$f_{D^*_s}/f_{D^*}$ & $1.10\pm0.02$ &\bridge \cite{fdsd} \\ 
%%\end{tabular*}
%%\end{ruledtabular}
%%\label{tab:input}
%%\end{center}
%%\end{table}

To obtain $\stwobg$, we minimize the $\chi^2$
\begin{equation}
\chi^2(2\beta + \gamma,\delta^{(*)},r^{(*)}) = \sum_{i}\!
\bigg(\frac{\tilde{x}_i-x_i}{\sigma_i}\bigg)^2 + \Delta(r^{(*)})\,,
\label{eq:chi2}
\end{equation}
where  $x_i = a, a^*, c_{\rm lep}, c^*_{\rm lep}$ are functions of the physics 
parameters (Eq.~\ref{acdep}), and $\tilde{x}_i$ are the corresponding measured values.
$\Delta(r^{(*)})$ is a continuous function that is set equal to $0$ within $30\%$ of the estimated $r^{(*)}$ 
(Eq.~\ref{lambdameas}), and is an offset quadratic outside this range, with the errors in Eq.~\ref{lambdameas}.
The additional $30\%$ error attributed on $r^{(*)}$ is due to
the unknown theoretical uncertainty on the validity of the $SU(3)$ symmetry assumption
and to neglecting $W$-exchange contributions to $A(\Bz{\to} D^{(*)+}\pi^-)$.
This error estimate is consistent with the spread in $r^{(*)}$ obtained using a variety of theoretical models~\cite{scr}.
The  $\sigma_i$ are the quadratic sums of the statistical and systematic uncertainties in Eq.~\ref{acmeas}.
Correlations between the $\tilde{x}_i$, at most $28\%$, have negligible influence on the results of this analysis.
The simultaneous analysis of two $B$ decay modes allows one to 
extract $|\stwobg|$. % independently of $|\cos\delta^{(*)}|$.
%, and gives a measurement of $\gamma$ with a four-fold ambiguity.

\begin{figure}[!tbp]
\begin{center}
\epsfig{figure=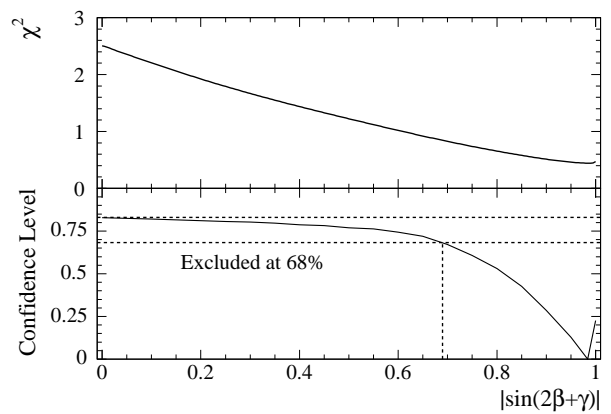,width=0.92\linewidth,clip=}
\end{center}
\caption{Dependence of  $\chi^2$ on $|\sin(2\beta+\gamma)|$ (top) and of the frequentist confidence
  level of the agreement of the data with expectations
  as a function of the   hypothesis on
  $|\sin(2\beta+\gamma)|$  (bottom). The assumptions on $r$ and
  $r^*$ are contained in the definition of $\chi^2$ (Eq.~\ref{eq:chi2}).
  The dashed horizontal lines indicate the $68\%$ and $83\%$ confidence levels
  (defined in the text).
}
\label{chi2}
\end{figure}

Figure~\ref{chi2} shows the minimum $\chi^2$ for each value of $|\stwobg|$. The absolute minimum occurs for 
$|\stwobg|=0.98$, where $\rm \chi_{ min}^2/d.o.f.=0.44/1$. 
The values of $r^{(*)}$ that minimize the $\chi^2$ are consistent with the input values within their statistical errors.
Because of the large uncertainties on the fit parameters and their limited physical range, 
the $\chi^2$ curve is non-parabolic. 
Thus to obtain a probabilistic interpretation to the results, we 
consider, for each of many values of $\sin(2\beta+\gamma)$, a large number of 
simulated experiments with the same characteristics as
the data.
We compute the consistency of the data with a given value of \stwobg\ by counting the fraction of simulated experiments 
for which $\chi^2(\stwobg)-\chi^2_{\rm min}$ is smaller than it is in the
data. 
This fraction, the frequentist confidence level, is shown in the lower
portion of Fig.~\ref{chi2}, from which we read that $|\stwobg|>\sbglim$ at $68\%$
C.L. 
We exclude the hypothesis of no \CP\ violation ($\stwobg=0$) at $83\%$ confidence level.
In order to study the impact of the assumed theoretical error on $r^{(*)}$, we doubled it to $60\%$
and we found that the lower limit on $|\stwobg|$ at $68\%$ C.L. drops from $0.69$ to $0.60$.

In conclusion, we studied the time-dependent \CP-violating 
asymmetries in fully reconstructed $\Bz{\to}D^{(*)\pm}\pi^{\mp}$ decays,
and measured the \CP-violating parameters listed in Eq.~\ref{acmeas}.
With some theoretical assumptions, we interpret the result in terms
of $\stwobg$ and we find that $|\stwobg|>\sbglim$ at $68\%$
C.L. and that $\stwobg=0$ is excluded at $83\%$ C.L.

We are grateful for the excellent luminosity and machine conditions
provided by our \pep2\ colleagues, 
and for the substantial dedicated effort from
the computing organizations that support \babar.
The collaborating institutions wish to thank 
SLAC for its support and kind hospitality. 
This work is supported by
DOE
and NSF (USA),
NSERC (Canada),
IHEP (China),
CEA and
CNRS-IN2P3
(France),
BMBF and DFG
(Germany),
INFN (Italy),
FOM (The Netherlands),
NFR (Norway),
MIST (Russia), and
PPARC (United Kingdom). 
Individuals have received support from the 
A.~P.~Sloan Foundation, 
Research Corporation,
and Alexander von Humboldt Foundation.


\begin{thebibliography}{99}
\bibitem{chconj}
Charge conjugation is implied in this Letter, unless 
otherwise stated. The superscript $(*)$ indicates that a symbol must be
considered both with and without the $*$ suffix.
\bibitem{note}
According to Ref.~\cite{fleischer}, the  strong 
phase for $B^0{\to} D^{*\pm} \pi^{\mp}$ is $\delta^* + \pi$.
\end{thebibliography}
\end{document}